\documentclass[traditabstract]{aa}
\usepackage{amsmath, amssymb, graphics, graphicx, hyperref, threeparttable}
\usepackage{natbib}
\bibpunct{(}{)}{;}{a}{}{,}

\begin{document}

\title{On the detection of point sources in CMB maps based on cleaned K-map} \titlerunning{On the detection of point sources in
  CMB maps based on cleaned K-map}

\author{H.G. Khachatryan\inst{1}$^{,}$\inst{2} \and G. Nurbaeva\inst{3} \and D. Pfenniger\inst{4} \and G. Meylan\inst{3}}

\institute{Center for Cosmology and Astrophysics, Alikhanyan National Science Laboratory, Alikhanyan brothers 2, Yerevan, Armenia
  \and Yerevan State University, A. Manoogian 1, Yerevan, Armenia \and Laboratoire d'astrophysique, Ecole Polytechnique F\'ed\'erale
  de Lausanne (EPFL), Observatoire de Sauverny, CH 1290 Versoix, Switzerland \and University of Geneva, Geneva Observatory, CH-1290
  Sauverny, Switzerland} \date{}

\abstract{We use the Wilkinson Microwave Anisotropy Probe 7-year data (WMAP7) to further probe point source detection technique in
  the sky maps of the cosmic microwave background (CMB) radiation. The method by Tegmark et al.\ for foreground reduced maps and the
  Kolmogorov parameter as the descriptor are adopted for the analysis of WMAP satellite CMB temperature data. Part of the detected
  points coincide with point sources already revealed by other methods. However, we have also found 2 source candidates for which
  still no counterparts are known, and identified 7 point sources listed in Planck Early Release Compact Source Catalogue as high
  reliability sources.}

\maketitle

\section{Introduction}

The cosmic microwave background (CMB) maps have been examined for detecting point sources as potential foregrounds, which could be
cosmological or Milky Way objects emitting thermally or non-thermally.  A catalog is made available on the
\href{http://lambda.gsfc.nasa.gov/product/map/current/}{NASA WMAP team web page}.  Various methods, including the wavelets and
needlets, have been used for the detection of point sources in pixelized sky maps (see, e.g., \citet{Sc,Bat}).  Most of the detected
sources in CMB maps coincide with known radio sources, quasars, blazars, although some sources are still unidentified
\citep{Wright,Jarosik2011}.

The Kolmogorov stochasticity parameter (KSP) has been already involved for such aims \citep{G2010}: some of the sources revealed by
that method initially had no counterparts, however their counterparts have been detected later by the Fermi satellite as gamma
sources and were included in its first source catalog 1FGSC. In the present study we continue the application of the KSP technique
to detect point sources in WMAP7 data set, and for that aim we borrow the maps cleaned by the method in \cite{Tegmark03}. While
applying the Kolmogorov method we analyze the mask issue, and we study the cross-correlations between the power spectra of the these
two types of maps, since the K-maps can carry information about the matter distribution in the Universe, particularly on voids
\citep{VG2009b}.
 
This paper is organized as follows. First we introduce briefly the Kolmogorov method, then we construct cleaned K-maps using the
Tegmark et al.\ method, modified for such maps.  We use the resulting maps to reveal the point sources, and to discuss the cross
power spectrum between the CMB temperature and the Kolmogorov maps (K-maps).

\section{Kolmogorov method and CMB maps}

In his original work, \citet{Kolm} introduced a method for determining whether the given random number sequence $X_i,\,{i=1,..,N}$,
ordered in an increasing way, obeys to a given statistic or not \citep{Arnold, Arnold_ICTP, Arnold_UMN, Arnold_MMS,Arnold_FA}. For
such a purpose the so-called empirical cumulative distribution function $F_N(x)$ is calculated,
\begin{eqnarray}
  F_N(x)=
  \begin{cases}
    0\ , & x < X_1\ \\
    k/N\ , & X_i \le x,\ \ k = 1,2,\ldots,N-1 \\
    1\ , & X_N \le x\,
  \end{cases}
\end{eqnarray}
where $k$ is the number of elements which obey to the relation $X_i\le x$. Having assumed a particular theoretical cumulative
distribution function (CDF) $F(x)$, the parameter $\lambda_N$ is easily calculated,
\begin{equation}
  \lambda_N = \sqrt{N}\, \sup_{x} \left| F_N(x)-F(x) \right|.
\end{equation}
Kolmogorov proved that in the limit $N\rightarrow\infty$, $\lambda_N$, which is a random variable, has a cumulative distribution
function $\Phi(\lambda)$ reading as
\begin{equation}
  \Phi(\lambda)=\sum_{j=-\infty}^{+\infty}{(-1)^{j}e^{-2j^{2}\lambda^{2}}}.
\label{distK}
\end{equation}
The function $\Phi(\lambda)$ can be expressed as a particular value of theta functions, since
$\Phi(\lambda)=\vartheta_{4}(0,e^{-2{\lambda}^2})=\vartheta_{3}(0,-e^{-2{\lambda}^2})$ \citep[see][16.27]{Abramovitz}.

\begin{figure}[t]
\includegraphics[width=0.45\textwidth]{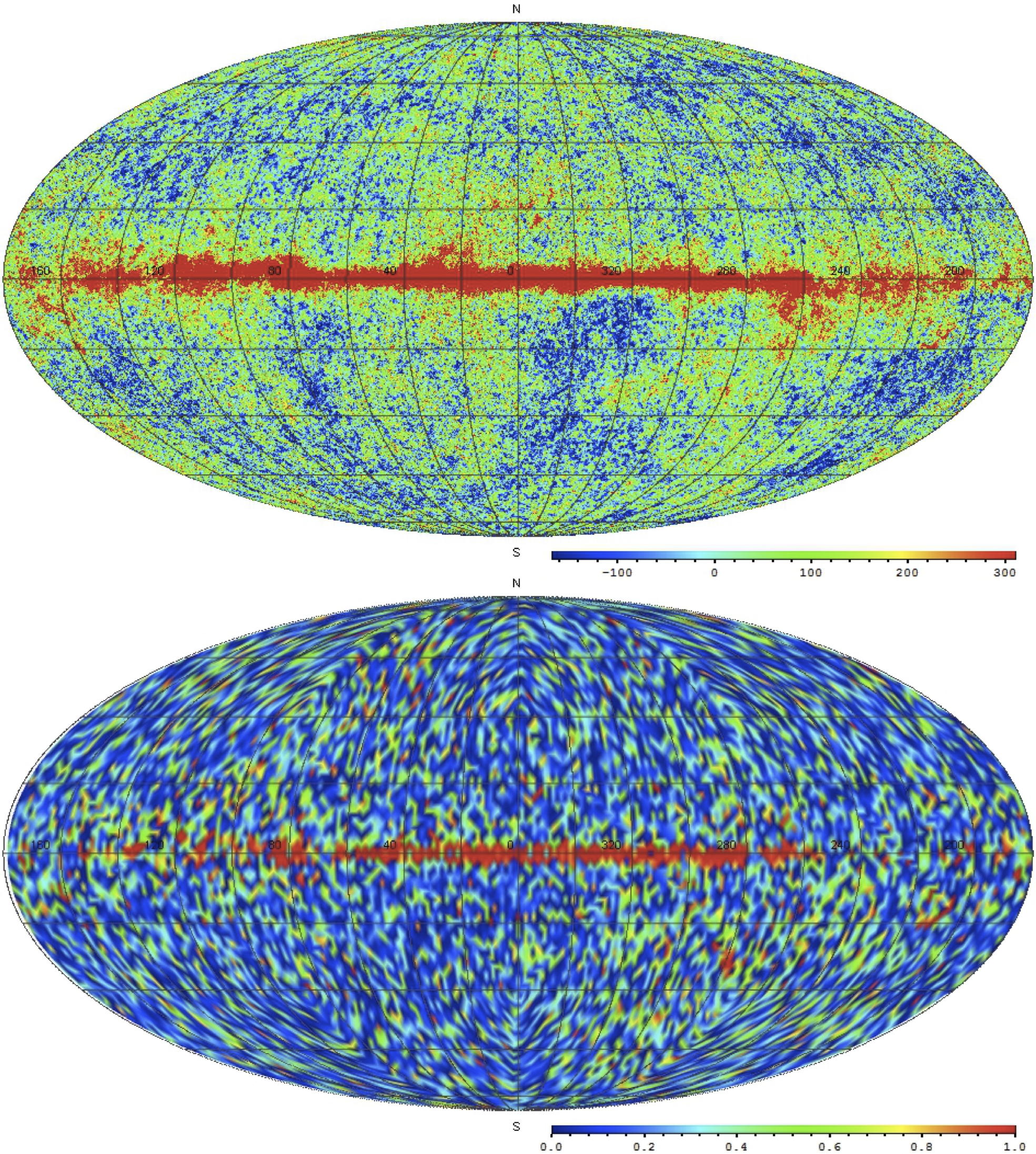}
\caption{WMAP7 Cosmic microwave background temperature (top) and Kolmogorov maps (bottom).}
\label{Maps}
\end{figure}

Some general worries were expressed by \citet{Frommert2012} for the use of the KSP parameter on the ground that the CMB 
fluctuations, despite Gaussian to a high degree, are angularly correlated.  In a short note Gurzadyan \& Kocharyan (2012)%
\footnote{arXiv:1109.2529}
counter-argued explaining that the KSP test should not be assimilated with the Kolmogorov-Smirnov goodness-of-fit test between two
different 1D distributions.  In this paper no distribution goodness-of-fit is performed, only the detection of abnormal pixels far
from a close to Gaussian distribution.  In this situation the eventual angular correlations affecting slightly the 1D Gaussian
distribution among pixels are irrelevant for abnormal pixel detection.

The CMB WMAP7 full sky map is Gaussian with high accuracy, but it also bears some regular signal like a noticeable angular auto-correlation at scales of about $1^\odot$. Do these correlations affect randomness at a given value of the CMB temperature on the sky, and is the KSP method applicable for it? \footnote{Note that the KSP method has been applied by Arnold in his original work with success to sequences very far from i.i.d. sequences, where he used simple completely deterministic sequences with arithmetic or geometric progression \cite{Arnold_FA,Arnold}, showing that the method can be applied to sequences including correlations.}  To check if the WMAP7 CMB map data values are independent random values, one can calculate the Pearson Chi-squared independence test on temperature values within randomly-chosen circles on the CMB sky with a radius of $2^\odot$. After applying the test we obtain that two random subsamples of this data are independent (the p-value is about 0.2).  Another aspect of angular autocorrelation in the CMB signal is that point sources add some subtle correlation into the observed WMAP CMB signal (see \citet{Fowler2010} as a good example of influence on angular autocorrelation power spectrum by point sources at high multipoles) as a steepening of the autocorrelation power spectrum. Thus angular autocorrelation cannot be removed from the CMB map without removing information about point sources. For more details see \cite{mod,VG2011}.

A known issue that can impact the KSP calculations is beaming and other instrumental correlation effects. In this paper we use only a single WMAP band CMB map for K-map calculations, therefore the beaming effect of different bands are not relevant. The mean value of the K-map varies very slightly between different bands if the Galactic disk region is not taken into account. Also the beaming effect adds some unexpected Gaussian noise into the CMB data. So it hardly has any effect on the K-map calculations aimed at detecting strong departure from a Gaussian distribution.

\section{Cleaned CMB K-map}

\subsection{WMAP7 W CMB and K-maps}

It is well known that the CMB sky has an nearly Gaussian distribution. So, the Gaussian distribution is used in Kolmogorov method to
construct a K-map. For every compact region of the CMB map containing 256 neighboring temperature pixels, one corresponding pixel
value of the K-map is obtained. For WMAP7 W band CMB map with $n_{\rm side}=512$ resolution parameter (details in \citet{Healpix}),
we obtain a K-map with $n_{\rm side}=32$ parameter. This means that the CMB map has $n_{\rm map}=12n_{\rm side}^2=3145728$ pixels and
the K-map $n_{\rm map}=12288$ pixels. This is due to the fact that KSP is a statistical parameter. Then the KSP distribution maps
over the whole sky can be obtained for WMAP7 data. It is seen that the Galactic disk region has higher and saturated KSP values,
which indicates that it has a non-Gaussian distribution, distinct from the CMB. Also a lot of pixels have a high value of KSP, but
most of them, as it will be shown below, are due to instrumental and other types of noise, also of non-Gaussian nature.
 
\begin{figure}[t]
\includegraphics[width=0.45\textwidth]{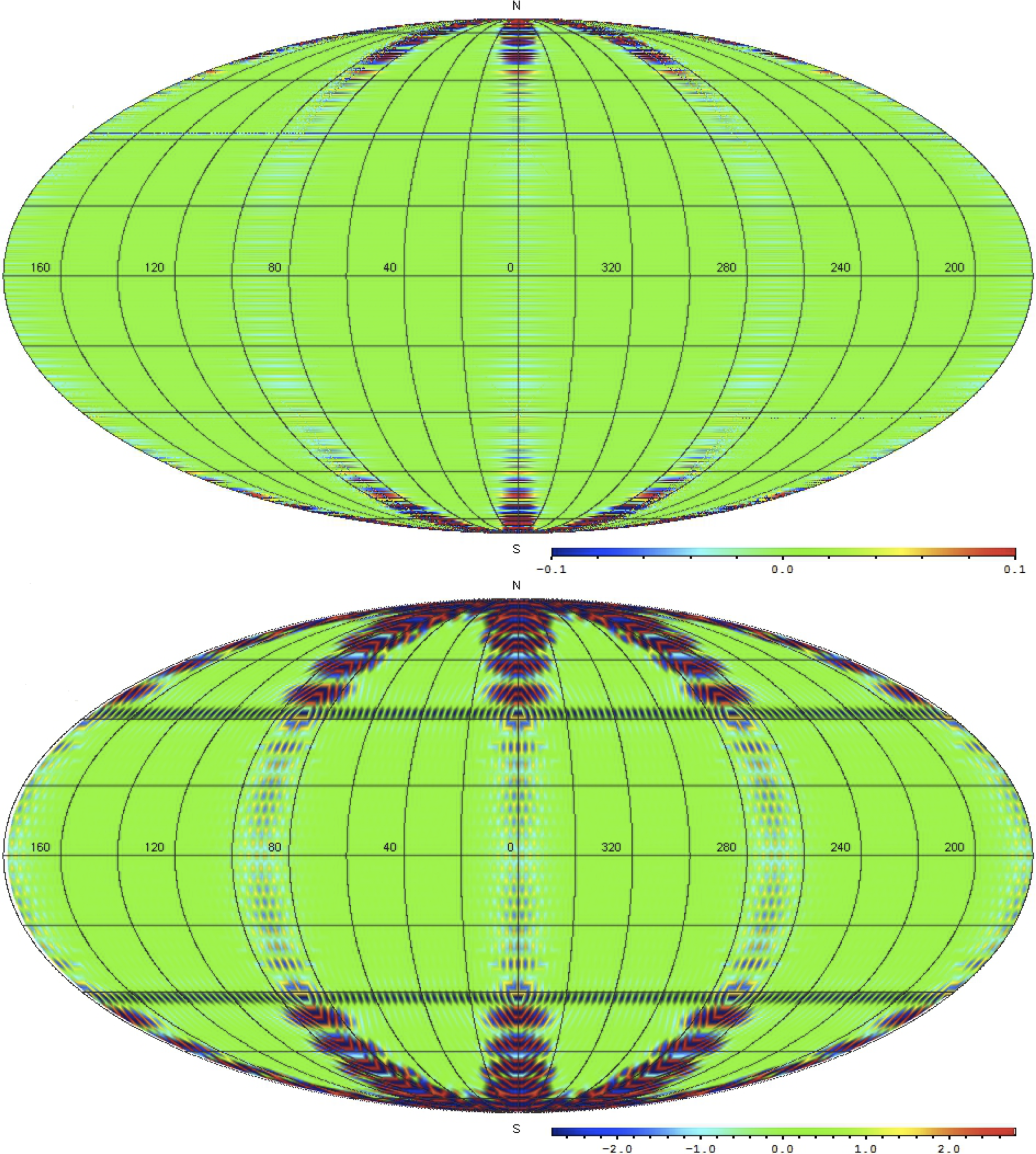}
\caption{HEALPIX error maps for $n_{side}=512$ (top) and $n_{side}=32$ (bottom) resolution parameters.}
\label{Errors}
\end{figure}

Throughout the analysis we mainly use the HEALPIX program package for calculation of the spherical harmonic coefficients and to
reconstruct the maps \citet{Healpix}. For some tedious manipulating procedures with the spherical harmonic coefficients $a_{lm}$
(see Eq.~(\ref{harm}) below) we use the GLESP program package \citet{GLESP}. For maps with low resolution (in our case for K-map,
$n_{\rm side}=32,n_{\rm map}=12288$) the $a_{lm}$ calculation and map reconstruction the use of HEALPIX program package led to 
accuracy problems. To test whether the calculation error is small or not for a low resolution map, we construct a unity
HEALPIX map with $n_{\rm side}=32$, and another one with $n_{\rm side}=512$, and run them through the $a_{lm}$ calculation and the map
reconstruction procedures. The result, multiplied by 1000, is given in Fig.~\ref{Errors}. For both cases this procedure adds some
non-isotropic noise which has almost zero mean ($e$) and a very small standard deviation ($\sigma$). We obtain $\langle
e_{512}\rangle= 10^{-6}$, $\left\langle e_{32}\right\rangle= 2\cdot 10^{-4}$, $\sigma_{512}=2.3\cdot 10^{-4}$,
$\sigma_{32}=3.71\cdot 10^{-3}$. Although some pixels around the poles have more error, the fraction of these pixels on the sky for
$n_{\rm side}=32$ is less than 0.5\%. So for $n_{\rm side}=32$, the error and standard deviation are sufficiently small, which
enables one to calculate the $a_{lm}$ and to construct the cleaned map for KSP using Tegmark et al.\ method.

\begin{figure}[tbp]
\centering
\includegraphics[width=0.45\textwidth]{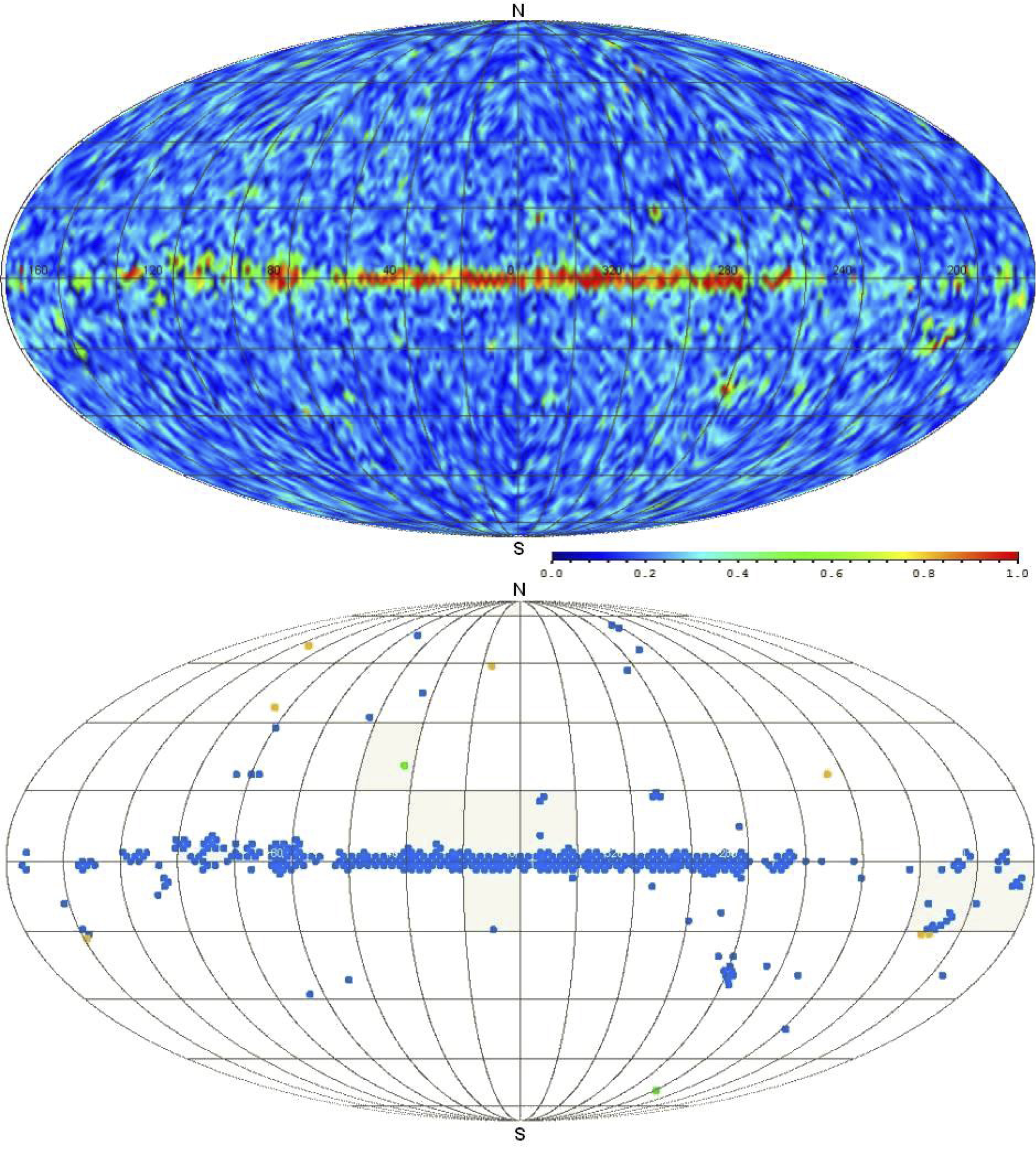}
\caption{Cleaned K-map (top) and regions with higher KSP value ($\Phi>0.5$) (bottom). The blue dots, outside Galactic and LMC
  regions, are point sources listed in the catalogs of \citet{Gold} and \citet{Lanz}, respectively. Seven yellow dots are point
  sources listed in the Planck Early Release Compact Source Catalogue \citep{Planck-sc}. The remaining two green dots are still
  unidentified possible point sources.}
\label{cKSP}
\end{figure} 

\subsection{Modified Tegmark et al. method}
Tegmark et al.\ method \citep{Tegmark03} is commonly used for obtaining a CMB foreground-reduced map from the original WMAP
CMB maps. In this method every map from different bands is weighted with weights $w_{l}^{i}$. Here $l$ is the multipole number and
$i$ the band index. It differs from the interlinear combination (ILC) method of weighting different bands suggested by the WMAP
team \citep{Jarosik2011} by dependence of weights on the multipole numbers. These weights are simply calculated from the
cross-correlation matrix between the bands.

\begin{figure}[t]
\centering
\includegraphics[width=0.45\textwidth]{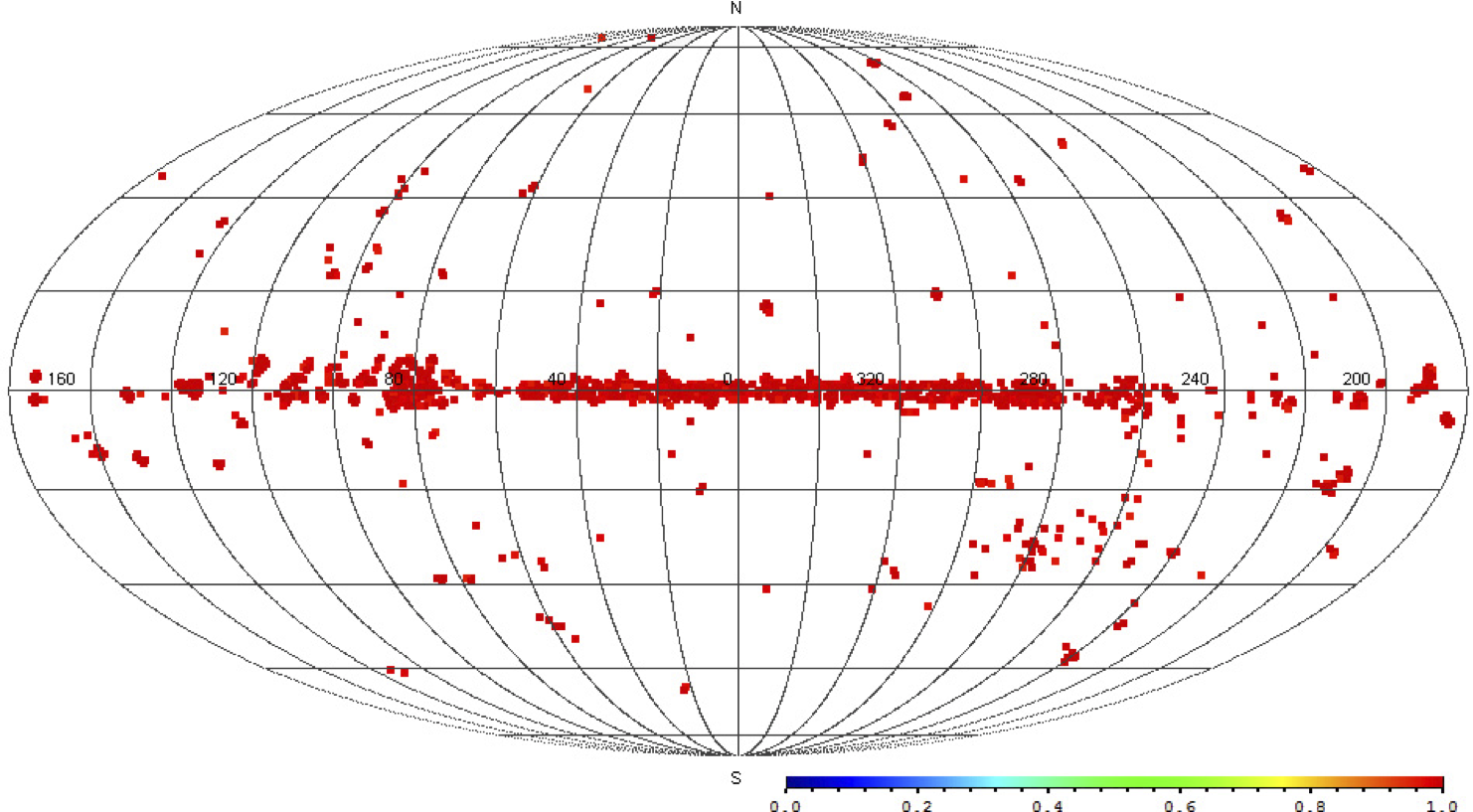}
\caption{Regions with higher KSP value ($\Phi>0.9$) of the K-map with $n_{\rm side}=64$ parameter. The first year Planck sattelite LFI CMB map with $n_{\rm side}=1024$ is used to obtain the K-map with $n_{\rm side}=64$. This is the first raw result derived from Planck data. Some more analysis should be done to extract point sources candidates from this data.}
\label{KSP64}
\end{figure}

We use Tegmark et al.\ method \citep{Tegmark03,Saha06,Saha08,VG2009b} to develop a cleaned CMB K-map for eight WMAP bands: Q1, Q2,
V1, V2, W1, W2, W3, W4. This method, based on the power spectrum comparison, assigns a weight $w_{l}^{i}$ for each map $i$ and
multipole $l$. In order not to distort the original map, $w_{l}^{i}$ should obey the relation
\begin{equation}
\sum_{i}{w_{l}^{i}}=1. 
\label{restriction}
\end{equation}
But a priori $w_{l}^{i}$ could be any real numbers, including negative ones. But KSP must belong to the interval $0 \leq \Phi \leq
1$, so the use of negative weights to construct the cleaned $a_{lm}$ and then to reconstruct the map could result in negative values
of $\Phi$ for some pixels. To avoid this problem, we normalize the weights by the following formula:
\begin{equation}
\acute{w}_{l}^{i}=\frac{w_{l}^{i}-w_{l}^\textrm{min}}{1-n\,w_{l}^\textrm{min}},
\label{weight}
\end{equation}
where $w_{l}^\textrm{min}$ is the minimal value of the original weights for fixed $l$ multipole. It is easy to see that for the new
weights $\acute{w}_{l}^i$ the relation $\sum_{i}{\acute{w}_{l}^{i}}=1$ holds.  This is the unique linear relationship satisfying
$\acute{w}_{l}^i\geq 0$ and minimally modifying the original $w_{l}^i$.

\begin{figure}[tbp]
\includegraphics[width=0.45\textwidth]{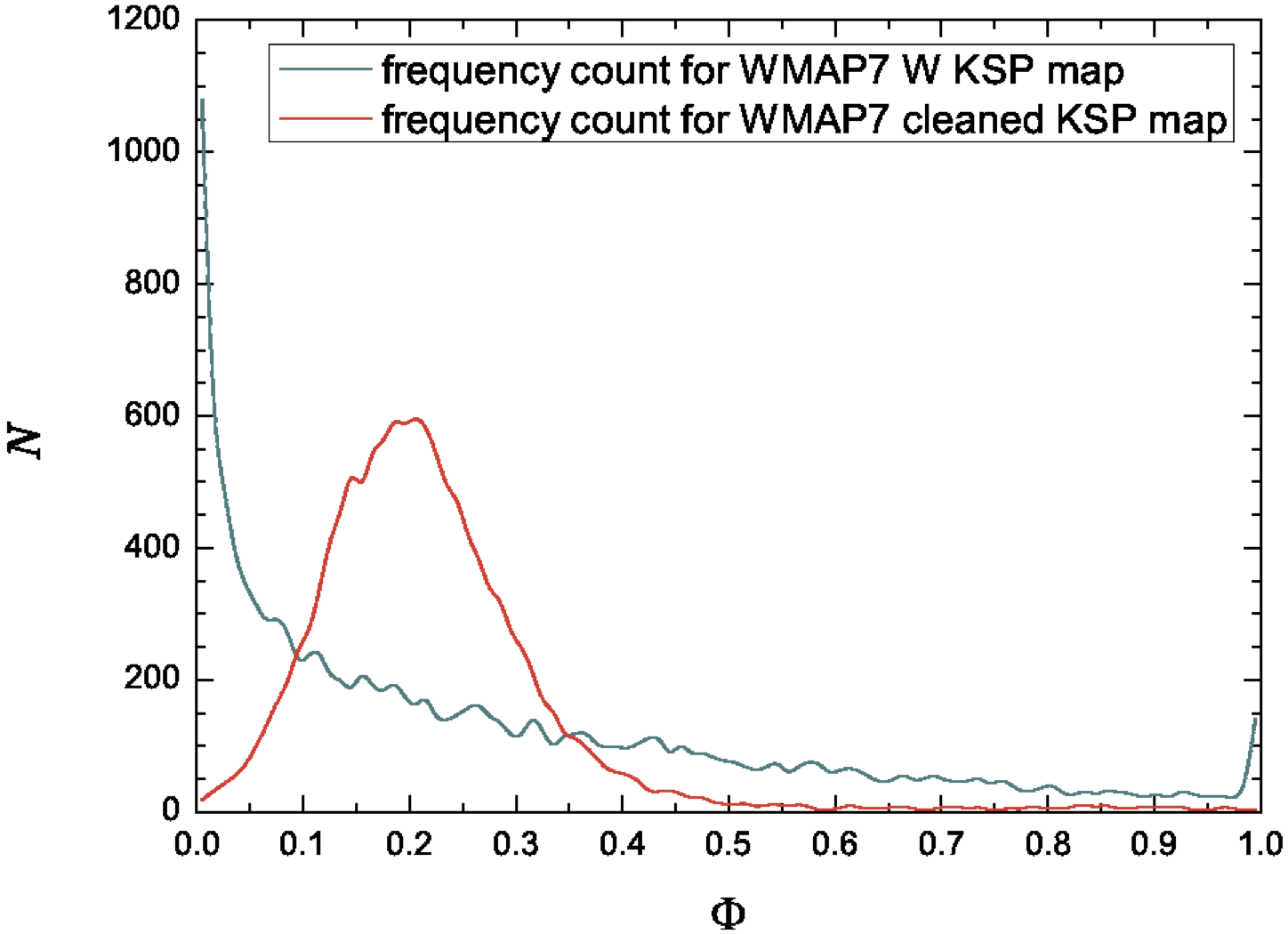}
\caption{Histograms for WMAP7 W band and cleaned K-maps.}
\label{freq}
\end{figure}

In the Kolmogorov method, if a sequence of random numbers $T_n$ obeys a theoretical distribution function $F(x)$ then $N$
realizations of this sequence gives $\Phi_{N}$, $\lambda_{N}$. The remarkable point of the method is that, $\Phi_{N}$ has a uniform
distribution and the mean value $\left\langle \Phi\right\rangle=0.5$. Therefore for a CMB K-map the value $0.5$ is a natural
threshold for distinguishing non-Gaussian areas from others, since Gaussian temperature pixels cannot have $\Phi> 0.5$. Further, the
cleaned K-map mean value and sigma are respectively $\left\langle \Phi\right\rangle=0.22$, $\sigma_{\Phi}=0.14$, so pixels with
$\Phi>0.5$ mostly exceed the 3-$\sigma$ region (see Fig.~\ref{freq}). In Fig.~\ref{cKSP} one can see the Galactic disk, the
Large Magellanic Cloud (LMC) and other possible point sources. There are in total 398 non-Gaussian pixels among 12'288 (about
$3.2\%$). Only 27 among them are outside the Galactic region ($|b|<20$) and the Large Magellanic Cloud ($l=280.4136$,
$b=32.9310$). So, these 27 regions are point source candidates. Indeed, most of them are found in the catalogs of \citet{Gold},
\citet{Lanz} and \citet{Planck-sc}. It is interesting that seven of the point sources are also listed in the Planck Early Release
Compact Source Catalogue as high reliability point sources \citep{Planck-sc}. Two possible point sources remain unidentified in any
of the known catalogs.

Another interesting feature of our cleaned K-map is that its distribution differs from the uncleaned maps in a minimal way. The
restrictions from the weighting scheme Eq.~(\ref{restriction}) keep the mean value almost unchanged. This means that structures in
the cleaned map are preserved during the cleaning procedure, but for a lot of regions the KSP value is decreased.

\subsection{Testing the method}

Here we represent a simulated Gaussian CMB map without any correlation and a K-map for it. We have not made any assumptions
  about the theoretical cumulative distribution function (i.e. mean and standard deviation of a Gaussian distribution), although it
  is fixed during the map generation. So we calculate these parameters for every region on the sky. This approach generates some
  false non-Gaussian regions with high values of the KSP due to varying these two parameters. As one can see, the K-map for the
  generated Gaussian map is blue almost everywhere, i.e., $\Phi$ is very small, but due to the numerical effect mentioned above,
  about $900$ of $12288$ pixels (about 7\%) of it are above the reasonable threshold of $0.5$. For different simulated maps this
  point appears at random positions on the maps. Very rare pixels have a KSP value higher than $0.9$. A similar situation appears in
  the real WMAP7 CMB maps. We can not assume any proper Gaussian distribution parameters for any region on the sky. Some remarkable
  non-Gaussian regions (such as the Galactic disk) can be seen in regions with high values of the KSP, but they are mixed with other
  regions affected by numerical inaccuracies. Hence, our problem is to distinguish false and real non-Gaussian regions (such as
  point sources) in the WMAP CMB map without considering in details numerical and statistical problems. For this purpose the Tegmark
  et al. method is used to clean the K-maps from different bands. Because both thresholds of KSP value (0.5 or 0.9) are acceptable
  here we use the lower one.

\begin{figure}
		\includegraphics[width=0.45\textwidth]{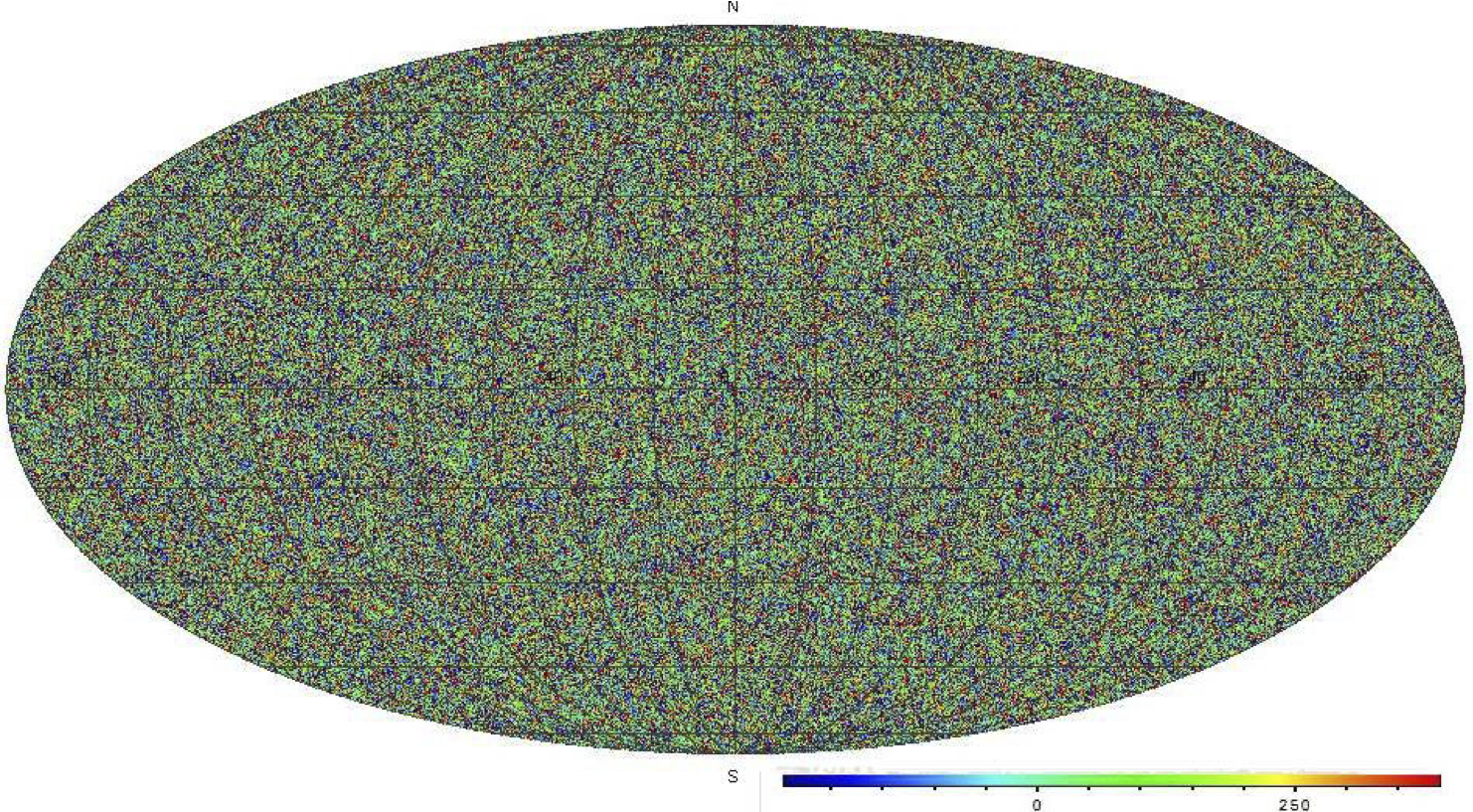}
	\label{fig:gauss_map}
	\caption{Simulated Gaussian CMB map without any correlation ($n_{side}=512$). Standard deviation and mean value are chosen from WMAP W band.}
\end{figure}

\begin{figure}
		\includegraphics[width=0.45\textwidth]{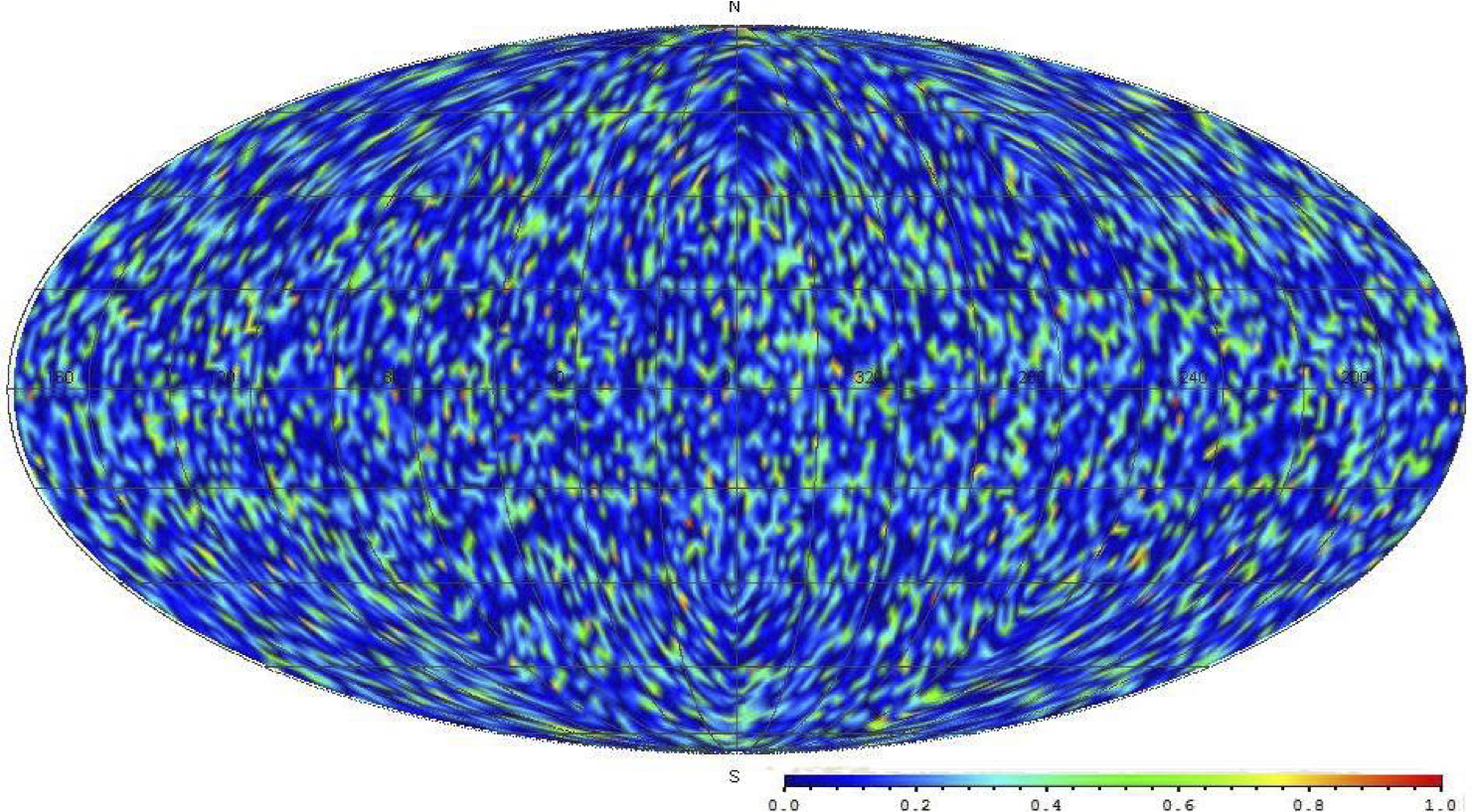}
	\label{fig:kmapgauss_map}
	\caption{K-map for the simulated Gaussian CMB map without any correlation($n_{side}=32$).}
\end{figure}

\begin{figure}
		\includegraphics[width=0.45\textwidth]{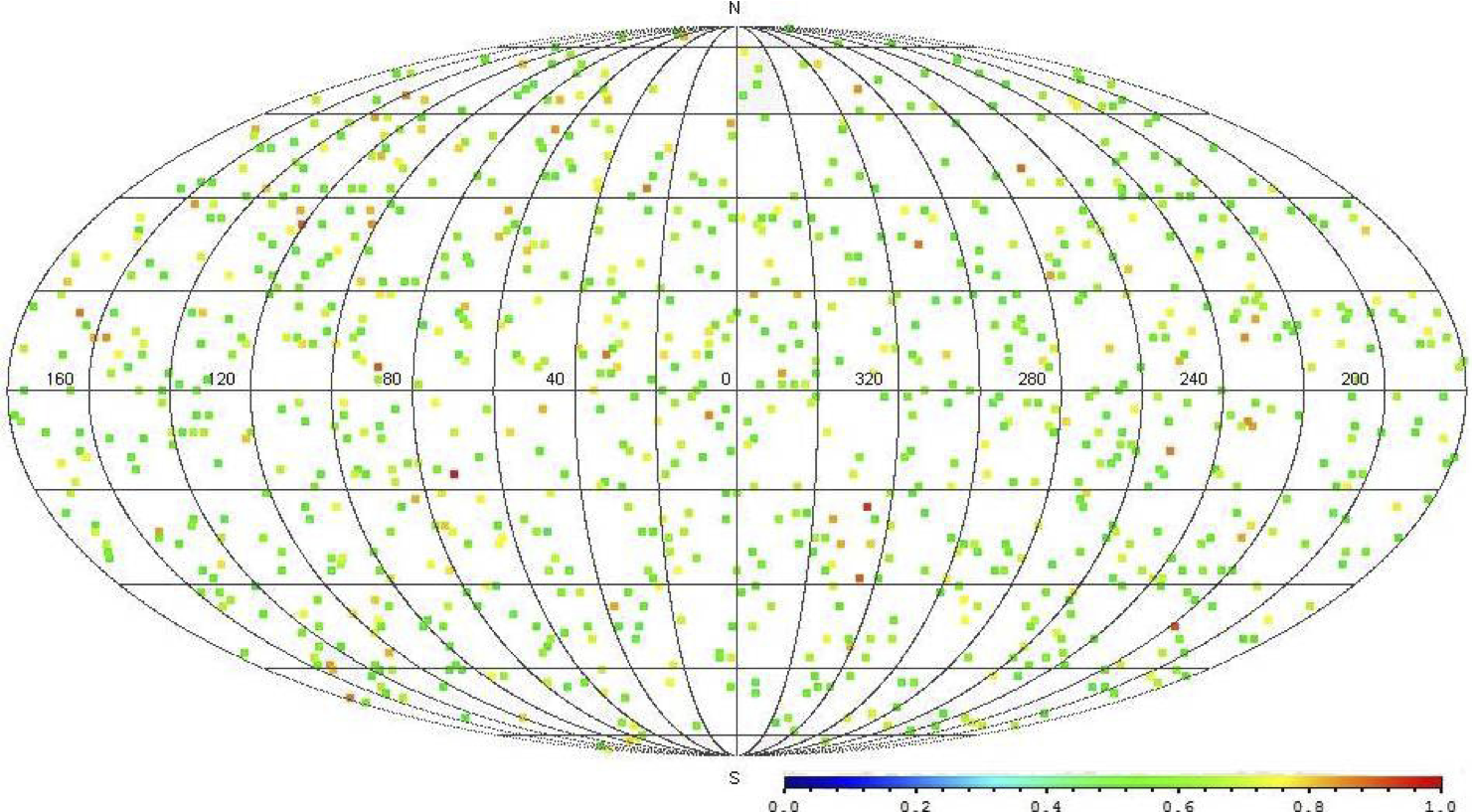}
	\label{fig:badgauss_map}
	\caption{Simulated Gaussian K-map points above 0.5 value.}
\end{figure}

\begin{figure}
		\includegraphics[width=0.45\textwidth]{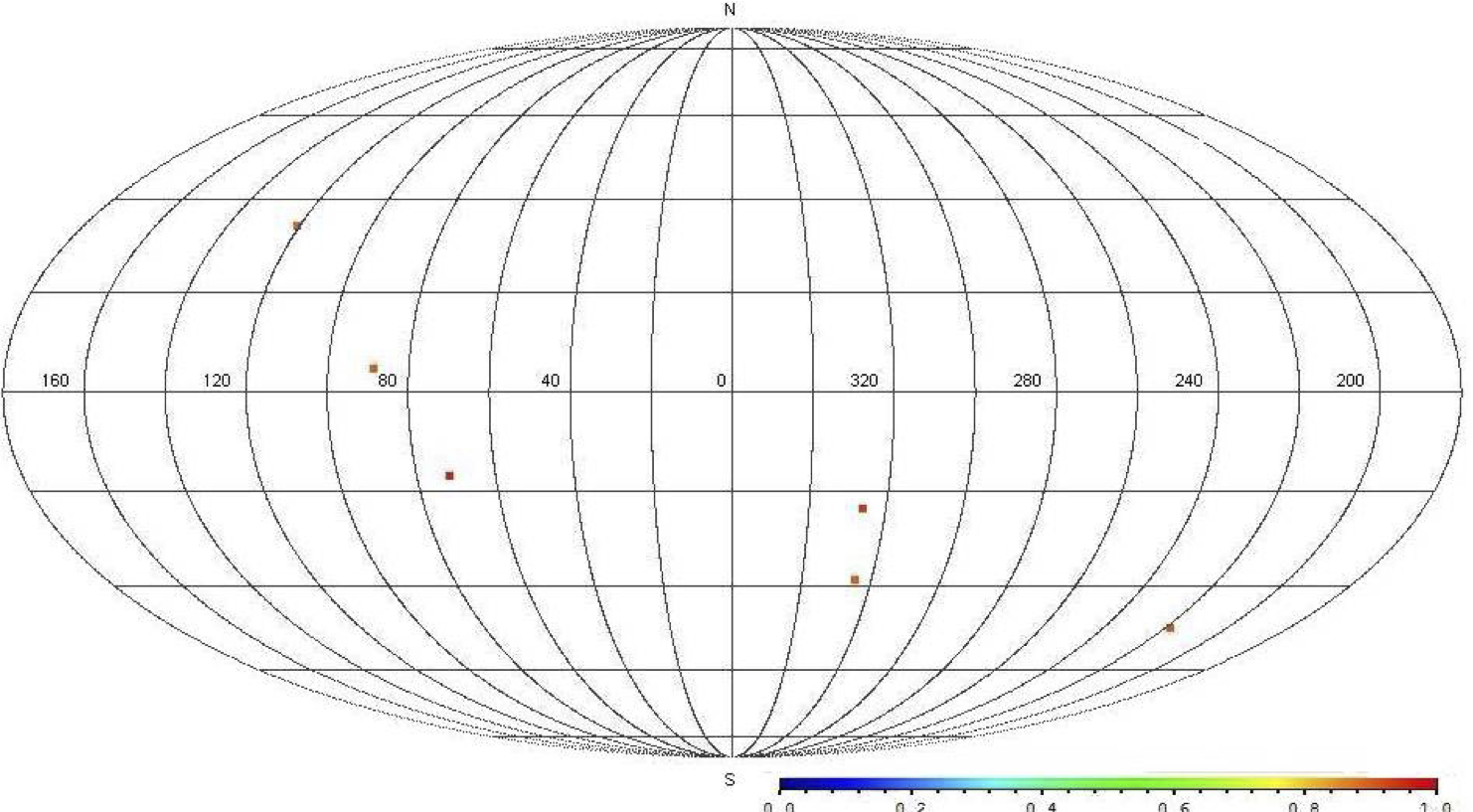}
	\label{fig:0.9gauss_map}
	\caption{Simulated Gaussian K-map points above 0.9 value.}
\end{figure}

\begin{figure}
	\centering
		\includegraphics[width=0.45\textwidth]{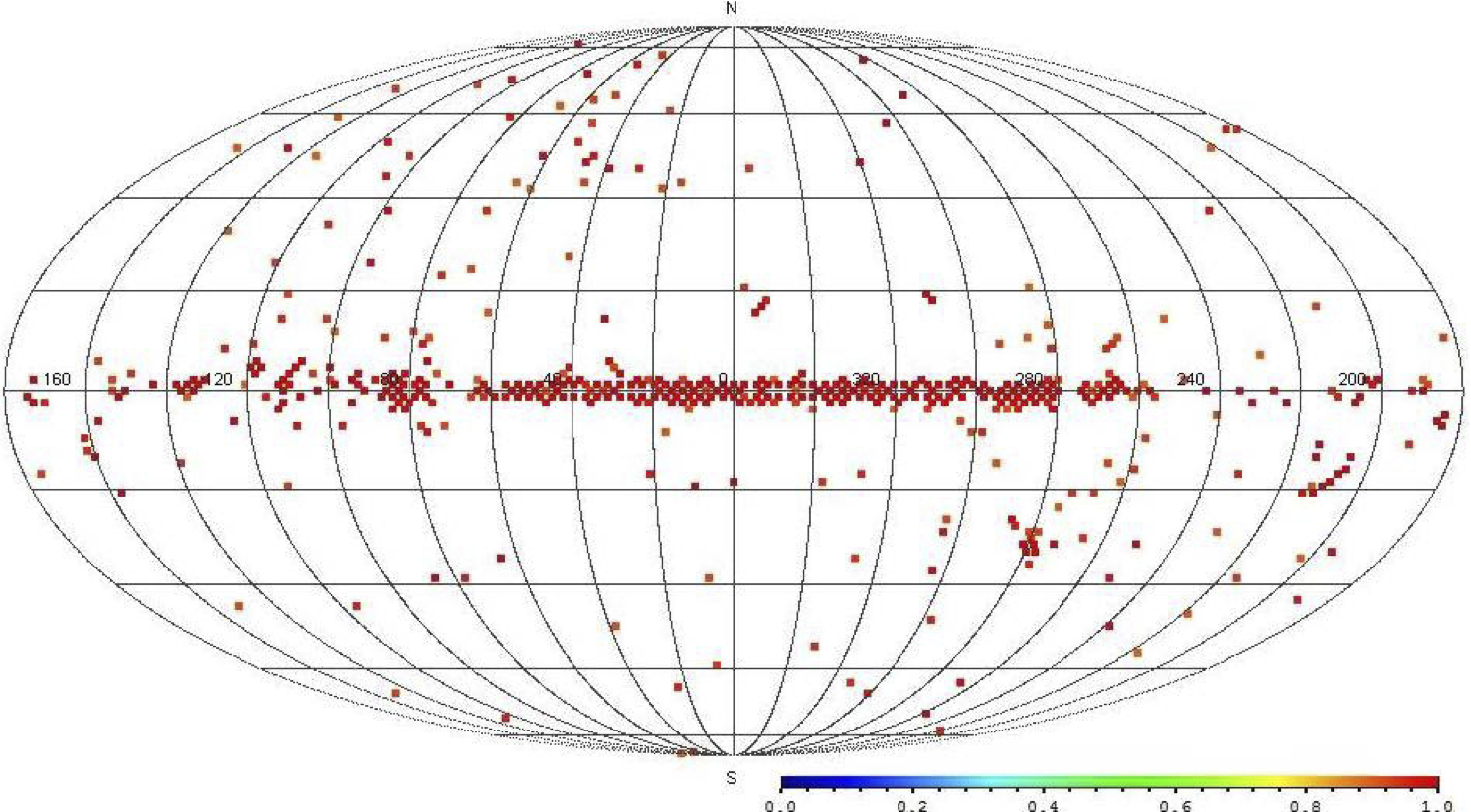}
	\label{fig:0.9cmp_map}
	\caption{CMB K-map points above 0.9 value.}
\end{figure}

\begin{figure}[t]
\includegraphics[width=0.45\textwidth]{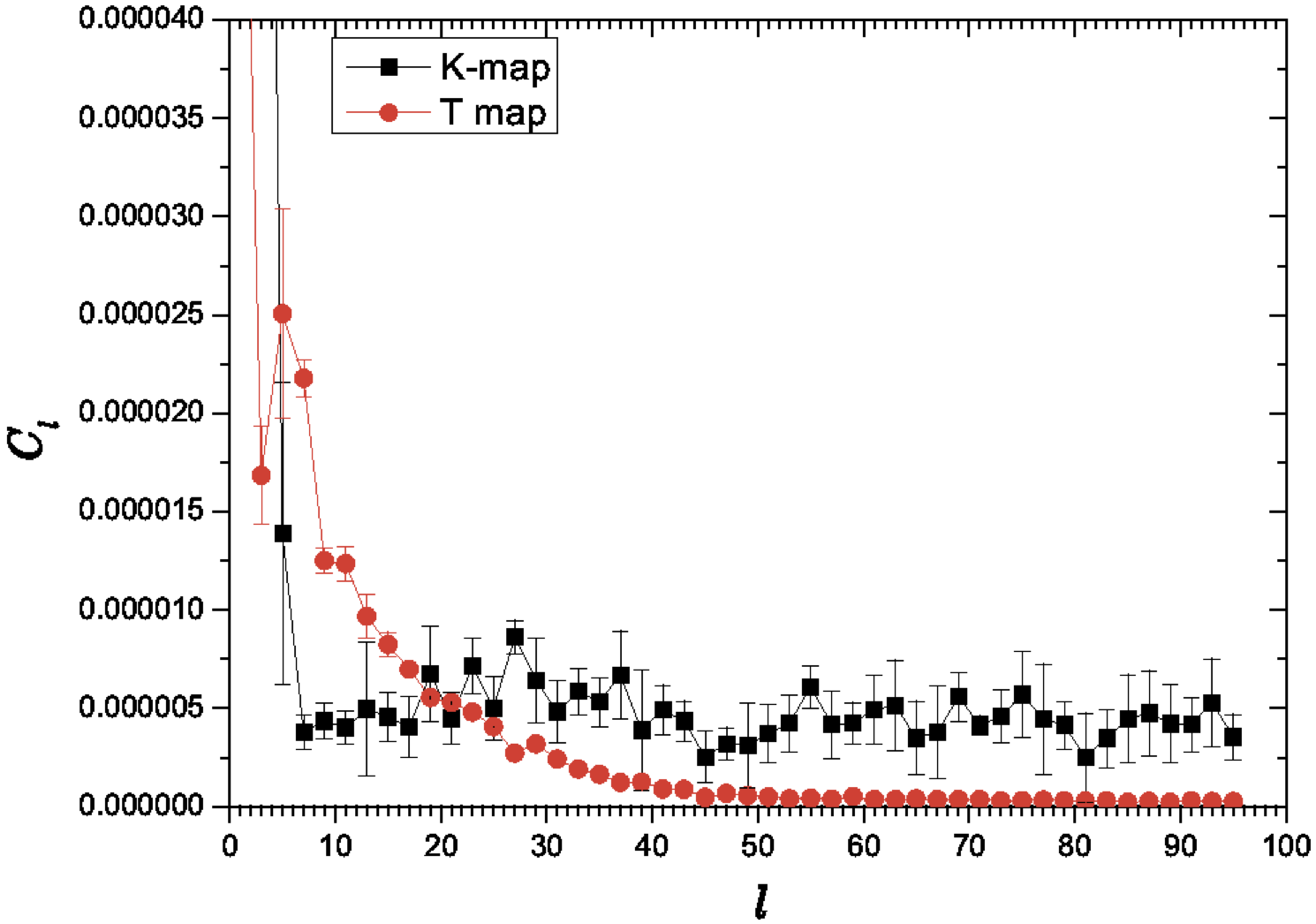}
\caption{CMB temperature and K-map correlation function power spectra.}
\label{Cl}
\end{figure}

\section{CMB and K-map power spectra}
Any full sky map can be represented via a series of Legendre spherical functions $Y_{lm}(\theta,\varphi)$ \citep{Hivon2002}, where 
\begin{eqnarray}
T(\theta ,\varphi )&=&\sum_{l,m}a_{lm}Y_{lm}(\theta ,\varphi ), \nonumber \\
a_{lm}&=&\int T(\theta ,\varphi ) \, Y^{\ast}_{lm}(\theta ,\varphi )\sin \theta \, d\theta \, d\varphi .
\label{harm}
\end{eqnarray}

As easily shown, the coefficients $C_l$ of the Legendre polynomials $P_l(\cos\theta)$ in the two point correlation function
$C(\theta)$ of the power spectrum are related to the $a_{lm}$ by
\begin{eqnarray} 
  C_{l}&=& \langle a_{lm}^{\ast}a_{lm}\rangle , \nonumber \\
  C(\theta )&=&\frac{1}{4\pi }\sum_{l,m}\left(2l+1\right) C_{l}\,P_{l}(\cos \theta ).
\label{Clr}    
\end{eqnarray}
To see if any correlation exists between the CMB temperature and the KSP parameter map, we degrade the resolution of the CMB
temperature map to $n_{\rm side}=32$ and also normalize it to the same $0\leq\,T\,\leq\,1$ interval as $\Phi$. So, instead of the
original temperature we use a dimensionless parameter $T$, to prevent any discrepancy between temperature and Kolmogorov maps. For
power spectrum and cross power spectrum calculations we use WMAP7 \citep{Jarosik2011} eight different bands $T$ and the Kolmogorov
maps (Q1, Q2, V1, V2, W1, W2, W3, W4). Then, the foreground reduced map is obtained using \citet{Tegmark03} weighting technique for
triplets of different bands for the calculation of the power spectra both for the temperature and the KSP correlation functions, via
standard technique as shown in \citet{Hinshaw03} and \citet{VG2009a}. The lines in Fig.~\ref{Cl} are the mean value of the $C_l$ for
different cross power spectra between different bands. The error bars are obtained as the variance square root of the different
cross power spectra.

The results for the foreground cleaned maps are shown in Fig.~\ref{Cl}. It can be seen that the $C_l$ have a different behavior
and the main striking fact is that they are crossed at $l\approx25$.

\section{Cross power spectra of CMB temperature and K-maps}

For cross power spectra calculation we use the common technique described in \citet{Hinshaw03} via taking the cross-correlation
power spectrum coefficients for different type of $a_{lm}$, as
\begin{equation}
\tilde{C}_l=\langle a_{lm}^{{\ast}T}a_{lm}^{\Phi}\rangle.
\end{equation}
 
Here we again use the foreground reduced maps \citep[see][]{Tegmark03} for the calculation of the cross power spectra between $T$
and $\Phi$, since the original ones are very noisy, while the aim is to get the smallest possible error bars in the final power
spectrum. The power spectrum estimation is done without taking into account the Galactic disk plane region, i.e., we use the window
function which is zero for the region $\theta=\pm20$ for both $T$ and $\Phi$, and unity elsewhere. This cutting method influences
mostly the even $l$ inducing some unreasonable peaks (for example at $l = 2, 12, 22, \ldots$) around the window function power
spectrum peaks. One could use \citet{Peebles73} method to reduce this effect on even $l$ values and adjust odd $l$ values but then
one would have to calculate the power spectrum at least up to $l=250$. The low resolution map of $\Phi$ allows the accurate
estimation of the power spectrum up to $l=97$ which makes impossible the use of this method. Therefore we have to use only odd $l$
values.

\begin{figure}[t]
\includegraphics[width=0.45\textwidth]{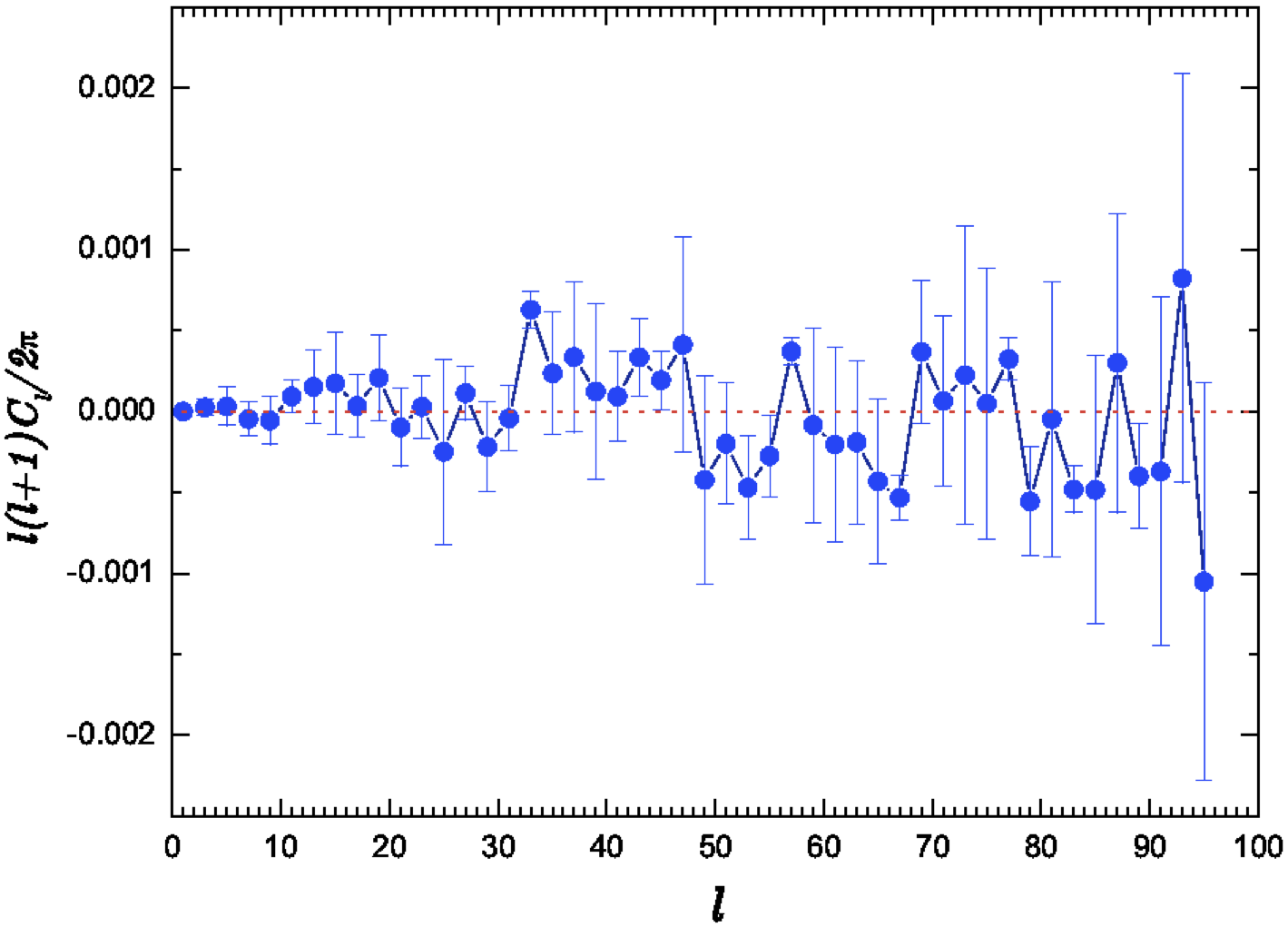}
\caption{Cross power spectrum of CMB WMAP7 $T$ and $\Phi$ maps.}
\label{Cross}
\end{figure}

In Fig.~\ref{Cross} no proper correlation can be seen between CMB WMAP7 $T$ and $\Phi$ maps.

\section{Conclusions}

Using the Kolmogorov stochasticity parameter for the detection of point sources in CMB maps, we proceeded from the K-maps cleaned
via the modified Tegmark et al.\,method. The novelty is that we applied this cleaning algorithm not to the usual temperature maps,
but to the K-maps. Other mask construction schemes are based on WMAP K and K1 band maps and are more affected by instrumental and
other types of noises. It appears that for about 85\% of the cleaned map, the Kolmogorov function has values in the interval
$0.0<\Phi<0.2$, which implies that the CMB maps are indeed Gaussian with high precision. The non-Gaussian pixels (with $\Phi>0.5$) are
rather rare ($398$), i.e., less than 4\%. Of course this result is derived for the cleaned K-map but it implies that different types
of noises add additional non-Gaussianities into the CMB maps, which may be analyzed by the methods in \citet{CMB-anomality} and
\citet{NGSIM}. These pixels, when outside the Galactic disk region $|b|>20$, indicate the positions of point source
candidates. While most of them have counterparts in existing catalogs, two of them are still unidentified.
 
Another type of non-Gaussianity discussed here is the fluctuations in the K-map. KSP is a statistical parameter so if one calculates
KSP for fixed numbers of elements, then it must have statistical fluctuations. But for the same reason, the full sky K-map should
have a uniform distribution, which is not so \citep{VG2011}. So in most cases these fluctuations have a non-Gaussian nature.

Some theoretical models \citep{CMB-fn,Spergel2000,Salopek90,Salopek91,Gangui94} predict primordial non-Gaussianities from inflation
era. This effect is very tiny but some authors tried to discover it through the CMB bi-spectrum. We then used a new method to
implement the cross-correlation between the CMB temperature and the K-map. Numerical modeling of such a problem was done in
\citet{mod}. It was shown that KSP is sensitive even to small departures from the theoretical distribution. Certain non-Gaussian
perturbations would appear in K-map as KSP perturbations. Also, if one uses a proper theoretical distribution function, no
correlation between the temperature and KSP should arise. Therefore in regions outside the Galactic disk certain correlations could
appear even in the presence of rather small non-Gaussianities. As one can see from Fig.~\ref{Cl} and Fig.~\ref{Cross}, any
correlation is rather difficult to see, although the intersection of the correlation functions of $T$ and $\Phi$ around $l=25$ might
be related to certain symmetries, see e.g.\ \citet{G_plane,VG2009a}.

\begin{threeparttable}
\begin{center}
\caption{Coordinates and ID-s of point sources detected in cleaned K-map.}
\label{ps}
\begin{tabular}{ccccc}
\scriptsize{l} & \scriptsize{b} & \scriptsize{KSP value} & \scriptsize{listed in} & \scriptsize{source ID} \\ \hline 
\scriptsize{15.00} & \scriptsize{58.92} & \scriptsize{0.50250} & \scriptsize{Planck} & \scriptsize{-}$^{b}$ \\ 
\scriptsize{43.59} & \scriptsize{27.28} & \scriptsize{0.50681} & \scriptsize{-} & \scriptsize{-}$^{a}$ \\ 
\scriptsize{45.00} & \scriptsize{49.70} & \scriptsize{0.50025} & \scriptsize{WMAP7} & \scriptsize{-}$^{c}$ \\ 
\scriptsize{63.28} & \scriptsize{41.81} & \scriptsize{0.57482} & \scriptsize{WMAP7} & \scriptsize{GB6 J1625+4134} \\ 
\scriptsize{67.50} & \scriptsize{-34.23} & \scriptsize{0.52509} & \scriptsize{Lanz et al.} & \scriptsize{GB6 J2148+0657} \\ 
\scriptsize{72.69} & \scriptsize{70.91} & \scriptsize{0.54416} & \scriptsize{WMAP7} & \scriptsize{GB6 J1419+3822} \\ 
\scriptsize{85.78} & \scriptsize{-38.68} & \scriptsize{0.83258} & \scriptsize{WMAP7} & \scriptsize{GB6 J2253+1608} \\ 
\scriptsize{97.03} & \scriptsize{24.62} & \scriptsize{0.56636} & \scriptsize{WMAP7} & \scriptsize{GB6 J1841+6718} \\ 
\scriptsize{99.84} & \scriptsize{38.68} & \scriptsize{0.51425} & \scriptsize{WMAP7} & \scriptsize{J1659+6827 G} \\ 
\scriptsize{99.84} & \scriptsize{24.62} & \scriptsize{0.56640} & \scriptsize{WMAP7} & \scriptsize{J1842+6808 QSO} \\ 
\scriptsize{105.47} & \scriptsize{24.62} & \scriptsize{0.52249} & \scriptsize{WMAP7} & \scriptsize{GB6 J1927+7357} \\ 
\scriptsize{106.50} & \scriptsize{44.99} & \scriptsize{0.61560} & \scriptsize{Planck} & \scriptsize{-}$^{b}$ \\ 
\scriptsize{132.19} & \scriptsize{66.44} & \scriptsize{0.50769} & \scriptsize{Planck} & \scriptsize{-}$^{b}$ \\ 
\scriptsize{157.50} & \scriptsize{-20.74} & \scriptsize{0.54468} & \scriptsize{Lanz et al.} & \scriptsize{GB6 J0336+3218} \\ 
\scriptsize{158.91} & \scriptsize{-22.02} & \scriptsize{0.54752} & \scriptsize{Planck} & \scriptsize{-}$^{b}$ \\ 
\scriptsize{195.47} & \scriptsize{-32.80} & \scriptsize{0.71974} & \scriptsize{WMAP7} & \scriptsize{PMN J0423-0120} \\ 
\scriptsize{210.94} & \scriptsize{-20.74} & \scriptsize{0.70129} & \scriptsize{Planck} & \scriptsize{-}$^{b}$ \\ 
\scriptsize{213.75} & \scriptsize{-20.74} & \scriptsize{0.62933} & \scriptsize{Planck} & \scriptsize{-}$^{b}$ \\ 
\scriptsize{238.33} & \scriptsize{-49.70} & \scriptsize{0.54351} & \scriptsize{WMAP7} & \scriptsize{PMN J0406-3826} \\ 
\scriptsize{246.09} & \scriptsize{24.62} & \scriptsize{0.51170} & \scriptsize{Planck} & \scriptsize{-}$^{b}$ \\ 
\scriptsize{251.72} & \scriptsize{-32.80} & \scriptsize{0.59206} & \scriptsize{WMAP7} & \scriptsize{PMN J0526-4830} \\ 
\scriptsize{258.75} & \scriptsize{-72.39} & \scriptsize{0.65437} & \scriptsize{-} & \scriptsize{-}$^{a}$ \\ 
\scriptsize{277.03} & \scriptsize{-35.69} & \scriptsize{0.50510} & \scriptsize{WMAP7} & \scriptsize{PMN J0537-6620} \\ 
\scriptsize{282.27} & \scriptsize{73.87} & \scriptsize{0.51290} & \scriptsize{WMAP7} & \scriptsize{GB6 J1230+1223}$^{c}$ \\ 
\scriptsize{283.50} & \scriptsize{75.34} & \scriptsize{0.73977} & \scriptsize{WMAP7} & \scriptsize{GB6 J1230+1223}$^{c}$ \\ 
\scriptsize{288.53} & \scriptsize{64.95} & \scriptsize{0.83234} & \scriptsize{WMAP7} & \scriptsize{QSO J1229+0203} \\ 
\scriptsize{304.77} & \scriptsize{57.40} & \scriptsize{0.78074} & \scriptsize{WMAP7} & \scriptsize{PMN J1256-0547}
\end{tabular}
\begin{tablenotes}
\item[a] \scriptsize{unidentified point sources}
\item[b] \scriptsize{possible point source identified in Planck Early Release Sources catalog \citet{Planck-sc}}
\item[c] \scriptsize{unidentified in WMAP7 catalog}
\end{tablenotes}
\end{center}
\end{threeparttable}

\begin{acknowledgements}
  We are grateful to V.\ Gurzadyan and colleagues in Center for Cosmology and Astrophysics for numerous comments and
  discussions. Many thanks also to O.\ Verkhodanov for information about GLESP.
\end{acknowledgements}

\bibliographystyle{aa}
\bibliography{K-mapa}
\end{document}